\documentclass{svproc}
\usepackage[utf8]{inputenc}
\usepackage{tikz}
\usepackage{graphics,multirow}
\usepackage{subcaption}
\usepackage{array}
\usepackage{subcaption}
\usepackage{xcolor}
\PassOptionsToPackage{hyphens}{url}
\usepackage{hyperref}
\usepackage{amsmath}
\usepackage{amsfonts}
\usepackage{amssymb}
\usepackage{mathtools}
 \hypersetup{
    colorlinks=true,       
    linkcolor=blue,          
    citecolor=blue,        
    filecolor=blue,         
    urlcolor=blue        
}
\definecolor{mygreen}{RGB}{0,214,0}

\newcommand\von{{\textsc{Von}}}
\newcommand\fon{{\textsc{Fon}}}
\newcommand\hon{{\textsc{Hon}}}
\newcommand\fston{$1^{\mathrm{st}}${\textsc{on}}}
\newcommand\nrep{N_{\mathrm{rep}}}
\newcommand\uprvec{\Pi_{\mathrm{Von}}^{\mathrm{U}}}
\newcommand\bprvec{\Pi_{\mathrm{1}}^{\mathrm{B}}}
\newcommand\hprvec{\Pi_{\mathrm{Von}}}
\newcommand\fprvec{\Pi_{\mathrm{1}}}
\newcommand\urank{K_{\mathrm{Von}}^{\mathrm{U}}}
\newcommand\brank{K_{\mathrm{1}}^{\mathrm{B}}}
\newcommand\hrank{K_{\mathrm{Von}}}
\newcommand\frank{K_{\mathrm{1}}}
\newcommand\vrank{K_{\mathrm{V}}}
\newcommand\reprank{K_{\mathrm{rep}}}

\begin{document}
\mainmatter
\title{PageRank computation for Higher-Order Networks}
\author{Célestin Coquidé \and Julie Queiros \and François Queyroi}
\authorrunning{Coquidé et al.}
\institute{Université de Nantes, LS2N, UMR CNRS 6004\\
44306 Nantes, France}
\maketitle
\begin{abstract}
Higher-order networks are efficient representations of sequential data. Unlike the classic first-order network approach, they capture indirect dependencies between items composing the input sequences by the use of \textit{memory-nodes}. We focus in this study on the variable-order network model introduced in \cite{xu16,saebi20}. 
Authors suggested that random-walk-based mining tools can be directly applied to these networks. We discuss the case of the PageRank measure. We show the existence of a bias due to the distribution of the number of representations of the items. We propose an adaptation of the PageRank model in order to correct it. Application on real-world data shows important differences in the achieved rankings. 
\end{abstract}

\section{Introduction}
\label{sec:intro}
\begin{figure}[ht]
    \centering    
    \begin{subfigure}{0.2\textwidth}
        \centering
        \caption{Input}
        \label{fig:ex_intro_input}
        \begin{tikzpicture}[scale = 0.5]
        \node[draw,text width=1.3cm] at(0,0){ 
            \texttt{y a c b c}\\
            \texttt{b c b c a}\\
            \texttt{x a c a}\\
            \texttt{y a c b}\\
            \texttt{x a c a c}\\
            \texttt{b c b}\\
            $\ldots$};
        \end{tikzpicture}          
    \end{subfigure}
    \begin{subfigure}{0.3\textwidth}
        \centering
        \caption{$P_{x}$ for $x \in \{xac,yac,bc\}$}  
        \label{fig:ex_intro_flow}
        \includegraphics[scale=0.15]{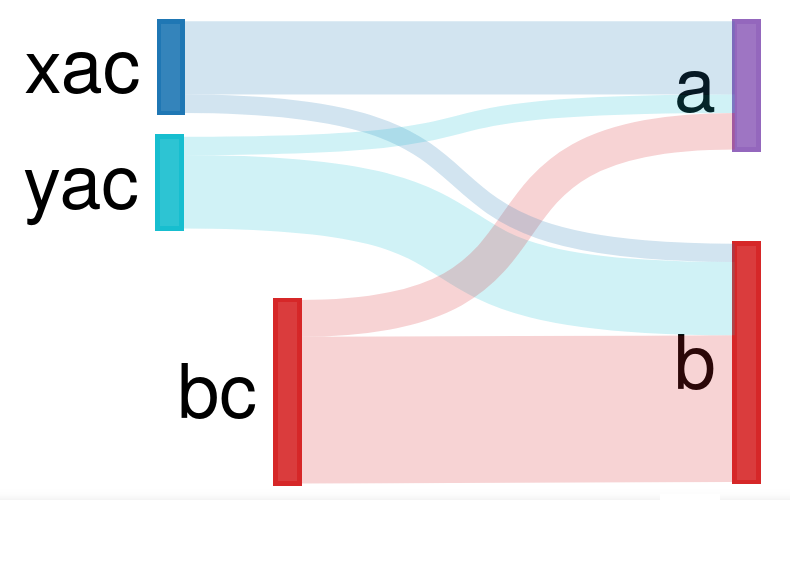}           
    \end{subfigure}
    \begin{subfigure}{0.3\textwidth}
        \centering
        \caption{\textsc{Von} network}
        \label{fig:ex_intro_hon}   
        \begin{tikzpicture}[scale = .95]
    \node[draw,scale=1.,circle] (X) at (-1,-1) {$x$};   
    \node[draw,scale=1.,circle] (Y) at (-1,1) {$y$};
    \node[draw,rectangle,scale=1.,fill=gray!20] (XA) at (0,-1) {$xa$};
    \node[draw,rectangle,scale=1.,fill=gray!20] (YA) at (0,1) {$ya$};
    \node[draw,scale=1.,circle] (A) at (0,0) {$a$};
    \node[draw,scale=1.,circle] (C) at (1,0) {$c$};
    \node[draw,rectangle,scale=1.,fill=gray!20] (XAC) at (1,-1) {$xac$};
    \node[draw,rectangle,scale=1.,fill=gray!20] (YAC) at (1,1) {$yac$};

    \node[draw,scale=1.,circle] (B) at (2,0)  {$b$};
    \node[draw,rectangle,scale=1.,fill=gray!20] (BC) at (2.,1) {$bc$};
    
    \draw[->,>=latex, line width=1.5pt] (X)--(XA);
    \draw[->,>=latex, line width=1.5pt] (Y)--(YA);
    \draw[->,>=latex, line width=1.5pt] (XA)--(XAC);
    \draw[->,>=latex, line width=1.5pt] (YA)--(YAC);
    \draw[->,>=latex, line width=0.1pt] (YAC)--(A);
    \draw[->,>=latex, line width=1.5pt] (YAC)--(B);
    \draw[->,>=latex, line width=1.5pt] (XAC)--(A);
    \draw[->,>=latex, line width=0.1pt] (XAC)--(B);
    \draw[->,>=latex, line width=1.5pt] (A) to [bend right=15] (C);
    \draw[->,>=latex, line width=1.pt] (C) to [bend right=5] (A);
    \draw[->,>=latex, line width=1.pt] (C)--(B);
    \draw[->,>=latex, line width=1.5pt] (B) to [bend right=10] (BC);
    \draw[->,>=latex, line width=1.5pt] (BC) to [bend right=10] (B);
    \draw[->,>=latex, line width=0.1pt] (BC)--(A);
    \end{tikzpicture}        
    \end{subfigure}
    \caption{Example of inputs and variable-order networks representing dependencies between \textit{items} $\Omega=\{x,y,a,b,c\}$. We can identify $2$nd or $3$rd order dependencies in \textbf{(b)}. For instance, when visiting $x$ then $a$ then $c$, the flow tends to return to $a$. The \textsc{Von}$(\mathcal{S})$ network in \textbf{(c)} only includes relevant dependencies. Subsequence $ac$ is not a relevant extension of the sequence $c$ since knowing we visited $a$ before $c$ does not impact the prediction of the next visited item.  Memory-nodes are displayed as grey rectangles. The set $\mathcal{V}(c)=\{xac,yac,bc,c\}$ are \textit{representations} of item  $c$.}
    \label{fig:ex_intro}
    \end{figure}
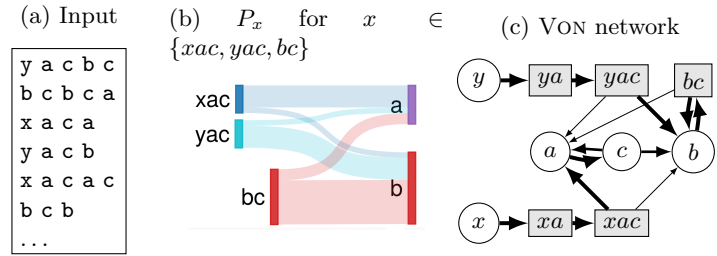

Network representation of real-world sequential data is an effective way to model complexity of interactions between items constituting them (flow of vessels between ports, city traffic, transfers between airports, \textit{etc.}).
A classic network model to represent sequential data is the pairwise interactions aggregation, extracted from the input data. This leads to a first-order Markov model of the sequences which can be represented by a first-order network (denoted \fston).
However, the input sequences might reveal higher-order dependencies between items (see Fig.~\ref{fig:ex_intro}). Recent works \cite{xu16,scholtes17} suggest that the {\fston} representation is not sufficient since it doesn't capture indirect dependencies in the underlying system.
Indeed, if such dependencies exist, a random walk performed on {\fston} may result in poor approximations of the flow of movements observed in the system.

Higher-order network ({\hon}) models are an alternative to the first-order Markov model approach.
In {\hon}, a node (or \textit{memory-node}) encodes a subsequence of varying length of items rather than a single item.
Most of the studied {\hon} are fixed-order networks (\fon$_{k}$) where the probability to reach the next item depends on the $k$ last visited ones \cite{scholtes17,rosvall14}.
Other studies consider variable-order models leading to variable-order network ({\von}) models \cite{xu16,saebi20}.
Random walks performed on {\hon} lead to better simulations of input sequences. One can therefore expect the results of random-walk-based network analysis tools such as  PageRank (PR)~\cite{page98}  to be more relevant.
In the context of {\von}, \cite{xu16} argue that such algorithms could be directly applied on variable-order network as they are still defined as weighted graphs. 
Nevertheless, authors have not investigated a possible bias in the resulting algorithm output due to the presence of memory-nodes.

In this study, we investigate the existence and nature of such bias when using the PR centrality measure.
We start by listing related works and discussing the ambiguity of the term \textit{higher-order} which is used for different purposes in the literature (Section~\ref{sec:rworks}). We then describe {\von}'s construction~\cite{saebi20} (Section~\ref{sec:von}).
In Section~\ref{sec:honpr}, we introduce the standard PR model and its direct application to {\von}.
We show that the teleportation mechanism used for computing PR values introduced a bias when applying it to {\von} networks. We introduce a correction as a bias-free PR model adapted for {\von}.
In order to assess the effect of the bias (and of our correction) in practice, we compare the various PR models on real-world sequential datasets (described in Section~\ref{sec:datasets}).
In section~\ref{sec:results}, we show that a direct application of PageRank on {\von} networks leads to an overestimation of the centrality of highly represented items. We also discuss the effect on ranking and its stability with a change of the damping factor parameter $\alpha$.
Finally, future works perspective are given in Section~\ref{sec:future_works}.
\section{Related Works}
\label{sec:rworks}

In the network science and data mining literature, the terms \textit{higher-order} or \textit{high-order} may relate to different concepts. For instance, in network clustering analysis, the term \textit{higher-order} used in \cite{yin17} refers to network motifs such as triangles and cycles. Authors designed a clustering algorithm preserving such motifs. In our study the term \textit{higher-order} refers as a network representation of a sequential data aiming to infer the indirect dependencies occurring in the sequences. In \cite{zhang20}, such dependencies were also considered, however, they are not been inferred from sequential data. 

Rosvall \textit{et al.}~\cite{rosvall14} introduced a fixed-order network of order 2 ({\fon}$_{2}$) and generalizations of PR and clustering algorithm to this new model. As an order of $2$ is limited for many real world applications, Scholtes~\cite{scholtes17} introduces to infer an optimal order leading to the most accurate fixed-order network. 
As PR computation may prove cumbersome on fixed-order networks, a {\hon} PR model called Multilinear PageRank was introduced in \cite{gleich15}. This model is based on a \textit{spacey surfer} whose next step doesn't depend on the previous one but on the most frequently visited ones.
We focus in this study on variable-order networks (\von) \cite{xu16,saebi20}. In this context, the PR computation of the items differs from the fixed-order model.

\section{Variable-order network representation}
\label{sec:von}
We detail in this section the method to build variable-order networks (\von) from input sequential data, as well as some important definitions and notations. We note $\mathcal{S}=\{s_1,s_2,\dots,s_l\}$ the set of $l$ sequences representing the input data. Each $s_{i} = \sigma^{1}_{i}\sigma^{2}_{i}\sigma^{3}_{i}\dots$ consists of a sequence of items (ports, airports, or any locations). The set of all items is denoted $\Omega$. 
The \textit{order} of a sequence $s$ denoted $|s|$ corresponds to its length. 
For two sequences $s_{1}$ and $s_{2}$, the sequence $s=s_{1}s_{2}$ is a concatenation of $(s_1,s_2)$ and we say that $s_{1}$ is a \textit{prefix} of $s$ and that $s_{2}$ is a \textit{suffix} of $s$.
For a sequence $s$, we call $c(s)$ the number of occurrences of $s$ in the dataset $\mathcal{S}$. 
The probability of finding item $\sigma$ next to sequence $s$ is estimated using relative frequencies
\begin{equation}
p(\sigma|s) = \frac{c(s\sigma)}{\sum_{\sigma ' \in \Omega}c(s\sigma ')}
\label{eq:transp}
\end{equation}
where $p(\sigma|s)$ is read "probability to find $\sigma$ having as context $s$". We define $p_s = \{p(\sigma|s), \sigma \in \Omega\}$ as the distribution of items following $s$.

\paragraph{Extraction of relevant extensions.} Memory-nodes are added according to \textit{relevant extensions} found in $\mathcal{S}$. The method we used was introduced in \cite{saebi20}. 
We say $s'$ is an \textit{extension} of $s$ if $|s'|>|s|$ and if $s$ is a suffix of $s'$. The extension $s'$ of $s$ is said to be relevant~\cite{saebi20} if
\begin{equation}
D_{KL} (p_{s'}||p_s) > \frac{|s'|}{\log_{2}(1+c(s'))}
\label{eq:dkl}
\end{equation}
where $D_{KL}$ denotes the \textit{Kullback-Leibler divergence}. Figure \ref{fig:ex_intro_flow} shows the distributions $p$ for the relevant extensions found in a toy example. The threshold used (right side of Eq.~\ref{eq:dkl}) make sit harder for longer and sparsely observed extensions to be found relevant.
The process used for relevant extensions extraction starts from the first-order sub-sequences. The condition in Eq.~\ref{eq:dkl} is recursively applied to extension of already detected extensions.
An upper-bound of $D_{KL}$ is used to stop the recursion. The construction of {\von} is therefore parameter-free.

\paragraph{Network construction.} {\von} is constructed in a way that a memory-less random walk performed on it is a good approximation of the input sequential data $\mathcal{S}$. This network is noted ${\von}=(\mathcal{V}, \mathcal{E}, w)$, where $\mathcal{V}$ is the set of nodes. These nodes represent all relevant extensions $s$ and all their prefixes (see Fig.~\ref{fig:ex_intro_hon}).
This ensures that any node $v$ representing a relevant extension of an item $\sigma$ is reachable during a random walk. 
We note  $\mathcal{V}(\sigma)$ the set of nodes representing item $\sigma \in \Omega$ 
 and $\nrep(\sigma)=|\mathcal{V}(\sigma)|$ the number of such representations.  

For each pair ($s,\sigma$), where $s$ is a relevant extension and $\sigma$ an item such that $p(\sigma|s)>0$, a directed link $s\rightarrow s^{*}\sigma$ is added to the network. The node $s^*\sigma$ represents the longest suffix of $s\sigma$ such that $s^*\sigma \in \mathcal{V}$. The weight of this link is $p(\sigma|s)$. 

\section{Application of PageRank to variable-order networks}
\label{sec:honpr}

In this section, we first introduce the PageRank (PR) measure and its direct application on {\von}. We discuss the effect of the distribution of $N_{rep}$ on PR probabilities distribution. 
In order to isolate this effect, we introduced a biased {\fston}PR model.
A bias-free model called Unbiased {\von} PR model is then introduced. 

\paragraph{Standard PageRank model ({\fston}PR).}
The PR measure is an efficient eigenvector centrality measure in the context of directed networks. It was implemented in Google's search engine by its inventors Brin and Page \cite{page98}. PR definition of node's importance can be interpreted as follows:  the more a node is pointed by important nodes, the more it is important. PR is equivalent to the steady state of a random surfer (RS) following a memory-less Markov process. The RS can follow links of the network with probability $\alpha$ or teleport uniformly towards a node of the network with probability $1-\alpha$ (it will also teleport from any sink node). These teleportations ensure that RS cannot be stuck in a sub-region of the network and that the steady state probability distribution is unique. The PR probability associated to the node $i$ is denoted $P(i)$. As item $i \in \Omega$ is represented by a single node $i$ in {\fston}, the PR probability associated to \textit{item} $i$ ($\Pi_1(i)$) is equal to $P(i)$. One can sort items by the decreasing order of their PR probabilities. We note $\frank$  
their ranks associated to $\Pi_1$ values.

\paragraph{Variable-order network PageRank ({\von}PR).}
In the case of {\von}s, the memory-less Markov process actually simulates the variable-order model as memory is indeed encoded into the nodes. 
Therefore, \cite{xu16} suggests that standard PR directly applied to {\von} will better reflect dependencies between items in the system than $\Pi_1$.
Since more than one nodes represent items in {\von}, 
\cite{xu16} defined the PageRank of an item as the probability for the RS to reach at least one of its representations (see Eq.~\ref{eq:mappedPR}). 
\begin{equation}
\Pi(i) = \sum_{\mathclap{v \in \mathcal{V}(i)}} P(v)
\label{eq:mappedPR}
\end{equation}
We denote by $\hprvec$ and $\hrank$ the PR values and ranking issued from {\von}PR model.

\begin{figure}
\centering
\begin{subfigure}[b]{0.4\textwidth}
  \centering
  \begin{tikzpicture}[scale = 1.]
  \node[draw,scale=1.,circle,fill=gray!25] (C) at  (0,0) {$c$};
  \node[draw,scale=.9,circle] (S1) at (4*360/10: 2.2cm) {$s_1$};
  \node[draw,scale=.9,circle] (S2) at (3*360/10: 2.2cm) {$s_2$};
  \node[draw,scale=.9,circle] (S3) at (2*360/10: 2.2cm) {$s_3$};
  \node[draw,scale=.9,circle] (S4) at (1*360/10: 2.2cm) {$s_4$};
  \node[draw=none,scale=1.]   (S5) at (5*360/10: 2.cm) {$\ldots$};
  \node[draw=none,scale=1.]   (S6) at (10*360/10: 2.cm) {$\ldots$};

  \draw[->,>=latex, thick] (S1) to [bend right=10] (C);  
  \draw[->,>=latex] (C) to [bend right=10] (S1);
  \draw[->,>=latex, thick] (S2) to [bend right=10] (C);  
  \draw[->,>=latex] (C) to [bend right=10] (S2);
  \draw[->,>=latex, thick] (S3) to [bend right=10] (C);  
  \draw[->,>=latex] (C) to [bend right=10] (S3);
  \draw[->,>=latex, thick] (S4) to [bend right=10] (C);  
  \draw[->,>=latex] (C) to [bend right=10] (S4);
  \draw[->,>=latex, thick] (S5) to [bend right=10] (C);  
  \draw[->,>=latex] (C) to [bend right=10] (S5);
  \draw[->,>=latex, thick] (S6) to [bend right=10] (C);  
  \draw[->,>=latex] (C) to [bend right=10] (S6);
  \end{tikzpicture}
  \caption{\textsc{Von} without round trips}
  \label{fig:toynet_1}
\end{subfigure}
\begin{subfigure}[b]{0.4\textwidth}
  \centering
  \begin{tikzpicture}[scale = 1.]
  \node[draw,scale=1.,circle,fill=gray!25] (C) at  (0,0) {$c$};
  \node[draw,scale=.9,circle] (S1) at (4*360/10: 1.2cm) {$s_1$};
  \node[draw,scale=.9,circle] (S2) at (3*360/10: 1.2cm) {$s_2$};
  \node[draw,scale=.9,circle] (S3) at (2*360/10: 1.2cm) {$s_3$};
  \node[draw,scale=.9,circle] (S4) at (1*360/10: 1.2cm) {$s_4$};
  \node[draw=none,scale=1.]   (S5) at (5*360/10: 1.2cm) {$\ldots$};
  \node[draw=none,scale=1.]   (S6) at (10*360/10: 1.2cm) {$\ldots$};
  
  \node[draw,scale=.8,circle,fill=gray!25] (S1c) at (4*360/10: 2.25cm) {$s_1c$};
  \node[draw,scale=.8,circle,fill=gray!25] (S2c) at (3*360/10: 2.25cm) {$s_2c$};
  \node[draw,scale=.8,circle,fill=gray!25] (S3c) at (2*360/10: 2.25cm) {$s_3c$};
  \node[draw,scale=.8,circle,fill=gray!25] (S4c) at (1*360/10: 2.25cm) {$s_4c$};

  \draw[->,>=latex, thick] (S1c) to [bend right=10] (S1); 
  \draw[->,>=latex, thick] (S1) to [bend right=10] (S1c);   
  \draw[->,>=latex] (C) to (S1);
  \draw[->,>=latex, thick] (S2c) to [bend right=10] (S2);
  \draw[->,>=latex, thick] (S2) to [bend right=10] (S2c); 
  \draw[->,>=latex] (C) to (S2);
  \draw[->,>=latex, thick] (S3) to [bend right=10] (S3c); 
  \draw[->,>=latex, thick] (S3c) to [bend right=10] (S3);  
  \draw[->,>=latex] (C) to (S3);
  \draw[->,>=latex, thick] (S4) to [bend right=10] (S4c); 
  \draw[->,>=latex, thick] (S4c) to [bend right=10] (S4); 
  \draw[->,>=latex] (C) to (S4);
  \draw[->,>=latex] (C) to (S5);
  \draw[->,>=latex] (C) to (S6);
  \end{tikzpicture}
  \caption{\textsc{Von} with round trips}
  \label{fig:toynet_2}
\end{subfigure}

\caption{\label{fig:toynet} Example of $\textsc{Von}$ models of trajectories where all flows go through an item $c$. In (a), when leaving $c$, a traveler goes uniformly to any of the satellites $s_i$. In (b), a traveler coming from $s_i$ always goes back to $s_i$. 
}
\end{figure}
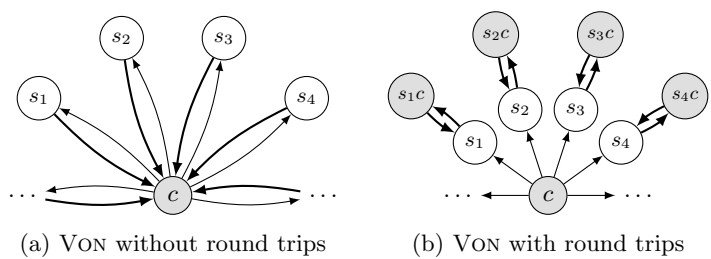

Since we use a random surfer, the more representations item $i$ has, the higher is the probability to teleport to one of them. As Eq.~\ref{eq:mappedPR} sums over representations of item $i$, this translates to a bias that is solely due to the teleportation mechanism.
We can illustrate this effect with a simple example (see Fig.~\ref{fig:toynet}). The value of $\hprvec (c)$ is always greater than or equal to $0.5$ in the situation illustrated in Fig.~\ref{fig:toynet_2} while it is always lower than or equal to $0.5$ in Fig.~\ref{fig:toynet_1}. Equality is achieved when $\alpha=1$ (i.e. when there is no teleportation). Although order $2$ dependencies exist in \ref{fig:toynet_2}, it is hard to justify why item $c$ should be ``more central'' in this case. 

\paragraph{($\nrep$)-biased PageRank model (Biased {\fston}PR)} In order to isolate the bias due to teleportations,  
we assume the transition probabilities associated to representation of item $i$ are all equal to $P_i$ \textit{i.e.} the representations of $i$ do not encode any different behaviour. 
This is equivalent to computing PR on {\fston} using a preferential teleportation vector $\mathbf{v}_{\mathrm{B}}$ depending on $\nrep$ as expressed in Eq.~\ref{eq:biasedprefvec}.

\begin{equation}
v_{\mathrm{B}}(j)=\frac{\nrep(j)}{\sum_{k \in \Omega} \nrep(k)}
\label{eq:biasedprefvec}
\end{equation}
The item PR values associated to this model and its resulting ranking are denoted $\bprvec$ and $\brank$ respectively.
Note that $\bprvec$ computed on \ref{fig:toynet_1} is equal to $\hprvec$ computed on \ref{fig:toynet_2} since the order-2 dependencies do not affect the centrality of $c$ in this example.

\paragraph{Unbiased {\von} PageRank model.}
In order to remove the bias discussed above, a modification of the teleportation vector is also used (see Eq.~\ref{eq:prefvec}).
Although several corrections are possible, the one chosen corresponds to the following random surfing process: teleportation is assumed to be the beginning of a new journey. It is therefore only possible to teleport uniformly to first-order nodes (see Eq.~\ref{eq:prefvec}).
\begin{equation}
v_{\mathrm{U}}(i)=
\left\{
\begin{array}{cl}
1/|\Omega | & \mbox{if $|i|=1$}\\
0 & \mbox{otherwise}
\end{array}
\right.
\label{eq:prefvec}
\end{equation}
It is easy to show that each node is reachable during the Markov process and therefore that the RS steady state is still unique using this teleportation vector.
The item PR values associated to this model and its resulting ranking are noted $\uprvec$ and $\urank$ respectively.

\section{Datasets and Experimental settings}
\label{sec:datasets}


\paragraph{Datasets.}
The three datasets used correspond to spatial trajectories. They differ however in terms of length, number of sequences, number of items, \textit{etc.}.
For each dataset and each sequence, we removed any repetition of items.
The code used and the datasets are available at \url{https://github.com/ccoquide/unbiased-von-pr/}.

\begin{itemize}
\item \textbf{Maritime :} Sequences of ports visited by shipping vessels, from April the 1st to July the 31st 2009.
Data are extracted from the Lloyd's Maritime Intelligence Unit. A variable-order network (\von) analysis of maritime is presented in \cite{xu16}.
\item \textbf{Airports :} US flight itineraries of the RITA TransStat 2014 database \cite{airport_url}, during the 1st quarter of 2011. Each sequence is related to a passenger, it describes passenger's trip in terms of airport stops. In \cite{scholtes17} and  \cite{rosvall14}, fixed-order network (\fon) representations of the data set are presented.
\item \textbf{Taxis :} Taxis rides into Porto City from July the 1st of 2013 to June the 30th of 2014. A sequence reports the succession of positions (recorded every 15 seconds) during a ride. The original data set \cite{taxi_url} was part of the ECML/PKDD challenge of 2015. Each GPS location composing the sequences is reported onto the nearest police station as it is suggested in \cite{saebi20}.
\end{itemize}
Sequences and networks statistics are reported in Table~\ref{tab:datasets}. 
We can observe that a large proportion of items have a large number of representations $\nrep$. The $\nrep$ values are far from being uniformly distributed. 

\begin{table}
\caption{Datasets and networks information.}
\label{tab:datasets}
\resizebox{\textwidth}{!}{%
\centering
\begin{tabular}{|l||r|r|r|r|r|r|r|r|}
\cline{1-9}
Dataset~& $|\mathcal{S}|$ ~& $|\Omega|$ ~& $|\mathcal{V}|$ ~& $|\mathcal{E}|$ ~& $max($order$)$ ~& Avg. $N_{\mathrm{rep}}$ ~& $Q_9(N_{\mathrm{rep}})$ & $\max(N_{\mathrm{rep}})$\\
\cline{1-9}
Maritime ~& 4K ~& 909 ~& 18K ~& 47K ~& 8 ~& 20 ~& 50 & 674\\
Airports ~&  2751K ~& 446 ~& 443K ~& 1292K ~& 6 ~& 995 ~& 1K & 34K\\
Taxis ~& 1514K ~& 41 ~& 4K ~& 15K ~& 14 ~& 99 ~& 250 & 382\\
\cline{1-9}
\end{tabular}
}
\end{table}

\paragraph{Experimental settings.}
For a given dataset, we compute PR values according to the different models with $\alpha = 0.85$ along with the corresponding rankings (see Section~\ref{sec:honpr}). Items having the same PR probabilities are ranked using the same order. 
In addition to the four models described in the previous section, we also report the following measures.
\begin{itemize}
\item \textbf{$\nrep$ ranking ($\reprank$)} is the ranking of items by decreasing order of $\nrep$. We quantify how $\nrep$-biased are other PR models by comparing them with this benchmark.
\item \textbf{Visit rank ($\vrank$)} is the ranking based on the probability of each item to occur in the input sequences. 
\end{itemize}
The visit rank is used as ``ground truth'' in \cite[Eq. 9]{scholtes17} for validation of the author's selection of fix-order model. 
However, we argue that this characterization is limited. 
For example, in the extreme situation where sequences are composed of only two items and can be viewed as a list of directed arcs, $\vrank$ would correspond to the ranking made from node degrees. More generally, using the item count as a centrality measure assumed an underlying symmetry in the system \textit{i.e.} every place is as much a destination as it is a departure. 



\section{Results}\label{sec:results}

We show here 
 that the bias effect is indeed important when looking at $\Pi$ values or the resulting rankings. Moreover, this is still true when 
 using alternative damping factor values.

\begin{figure}[h!]
\centering
\begin{subfigure}[b]{0.32\textwidth}
\centering
\includegraphics[width=\textwidth]{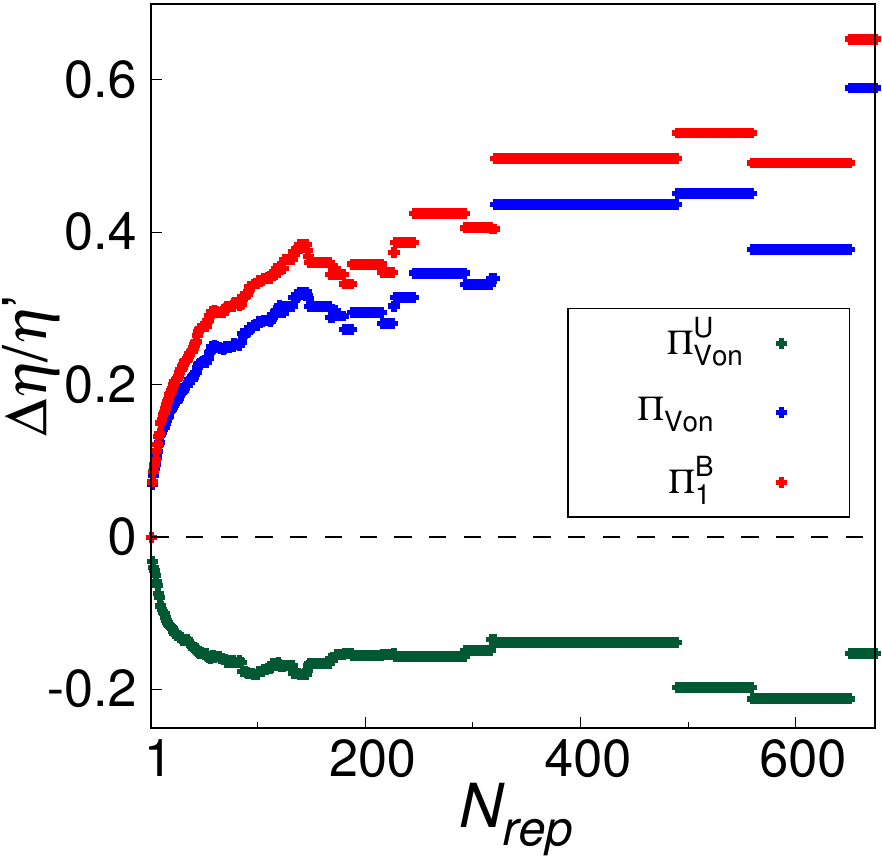}
\caption{Maritime}
\label{fig:deltaEta_maritime}
\end{subfigure}
\begin{subfigure}[b]{0.32\textwidth}
\centering
\includegraphics[width=\textwidth]{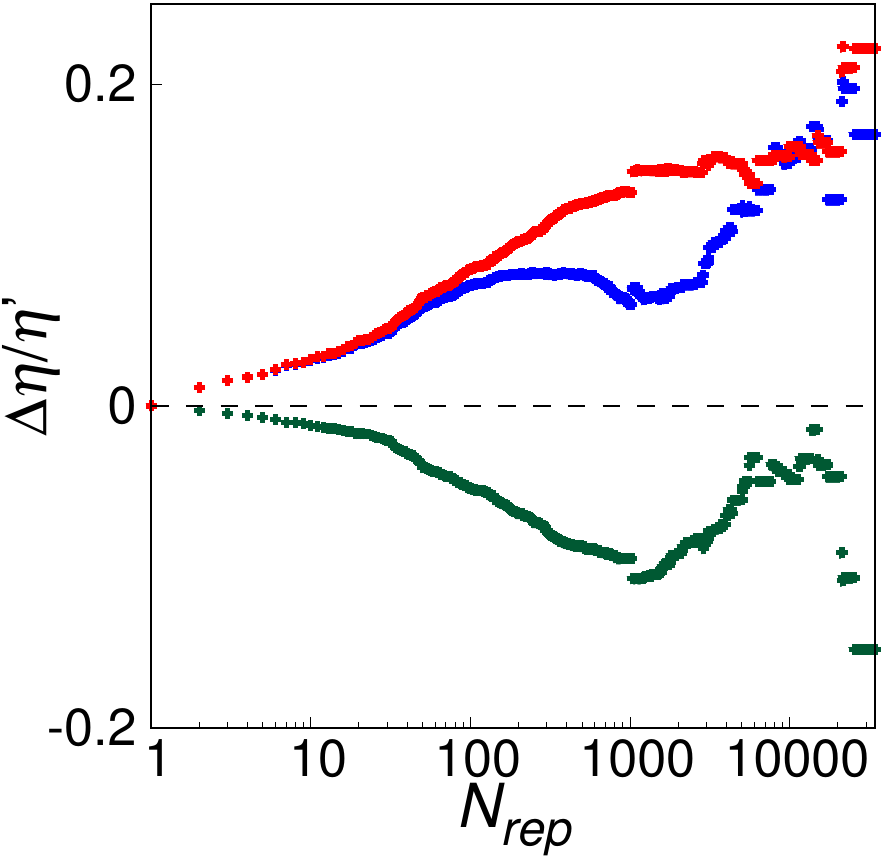}
\caption{Airports}
\label{fig:deltaEta_airports}
\end{subfigure}
\begin{subfigure}[b]{0.32\textwidth}
\centering
\includegraphics[width=\textwidth]{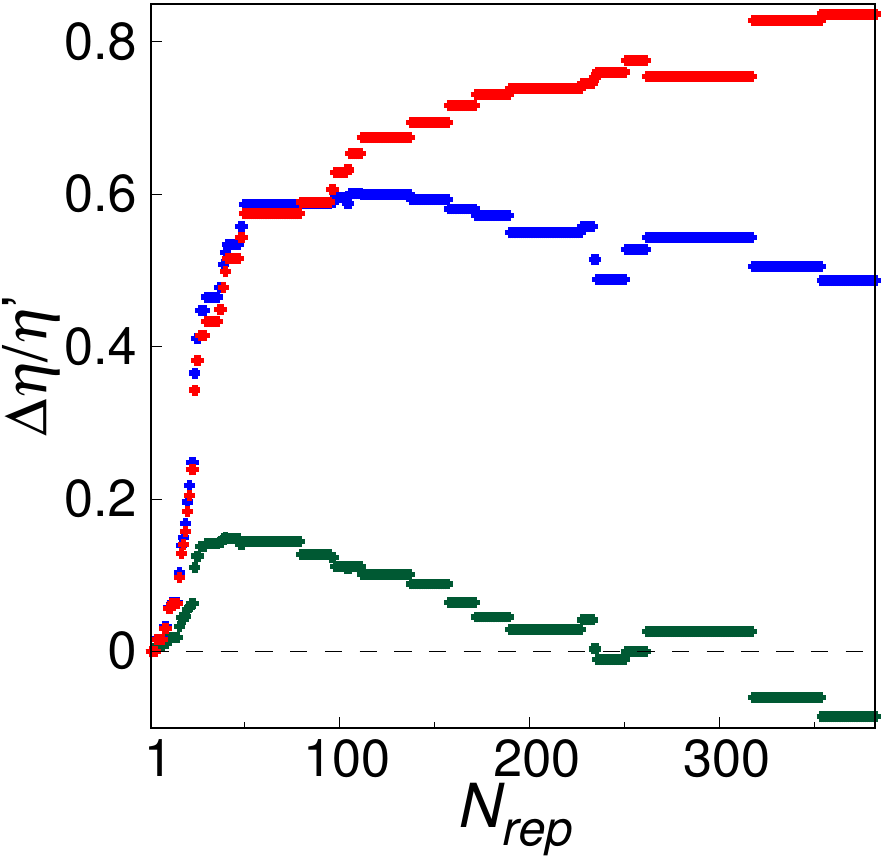}
\caption{Taxis}
\label{fig:deltaEta_taxis}
\end{subfigure}
\caption{Relative PageRank boost $\Delta \eta/\eta^{'}$ versus $\nrep$ for three PR models, with $\alpha=0.85$. For a given number of representations $\nrep$, $\eta(N_{\mathrm{rep}})$ is the probability to reach any item having at least $\nrep$ representations. $\eta'$ is related to {\fston}PR model. 
}
\label{fig:deltaEta}
\end{figure}

\paragraph{Evolution of $\Pi$ values with $\nrep$.}
We note $\eta(\nrep)$ the probability that a random surfer (RS) visits any item having at least $\nrep$ representations such as
\begin{equation}
\eta(\nrep) = \sum_{j\in \Omega}\Pi(j)\mbox{ with $\nrep(j) \geq \nrep$}
\label{eq:eta}
\end{equation}
The impact of $\nrep$ on PR probabilities is quantified by the relative PR boost $\Delta\eta/\eta^{'} = (\eta-\eta^{'})/\eta^{'}$ where $\eta^{'}$ is related to the {\fston}PR model.
We show the evolution of $\Delta \eta/\eta^{'}$ with $\nrep$ in Fig.~\ref{fig:deltaEta} for each dataset.
Both {\von}PR ($\hprvec$) and Biased {\fston}PR ($\bprvec$) models are the ones with the highest relative PR boosts. For example, in case of Maritime dataset (see Fig.~\ref{fig:deltaEta_maritime}), the relative PR boosts, at $\nrep=\nrep^{\max}$, equal to $60\%$ and $65\%$ respectively for these models (compared to {\fston}PR probabilities $\fprvec$). Moreover, we see that the PR boosts relative to $\hprvec$ fit well with $\bprvec$ ones.
The Unbiased {\von}PR probabilities ($\uprvec$) are impacted very differently. 
For the Airports and Taxis datasets, the distributions shape of boosts is similar to  {\von}PR's but with lower boost values.
Although relative PR boosts are the lowest for our model, the higher-order dependencies it encodes still lead to differences with {\fston}PR. 
However, the fact that PR probabilities are boosted for highly represented items doesn't necessarily lead to resulting biased PR rankings. Therefore, we investigate the changes in rankings when using the different models.

\paragraph{Rankings comparison.}
We quantified similarities between pairs of PR rankings by using both Spearman and Kendall correlation coefficients. Table~\ref{tab:correlations} displays similarities for each dataset. We observe high similarity between {\von}PR ($\hrank$) and Biased {\fston}PR ($\brank$) rankings. On the other hand, correlation coefficients between Unbiased {\von}PR rankings ($\urank$) and $\brank$ are lower. We also observe overall lower correlation coefficients with Taxis dataset which are probably due to the lower number of items.
Note that the visit rank ($\vrank$) is closer to $\brank$ or $\hrank$ (for Taxis). 
If we were to use $\vrank$ as a PR selection method, we would likely select our biased {\fston}PR model which does not include any higher order dependencies. This highlights the fact that $\vrank$ is not an efficient benchmark in the context of {\von}.

\begin{table}[h!]
\caption{
    Spearman ($r_{s}$) and Kendall ($r_{\tau}$) coefficients between PR rankings.
    }
\label{tab:correlations}
\centering
\resizebox{0.85\textwidth}{!}{%
\begin{tabular}{l|>{\centering\arraybackslash}p{1cm}>{\centering\arraybackslash}p{1cm}|>{\centering\arraybackslash}p{1cm}>{\centering\arraybackslash}p{1cm}|>{\centering\arraybackslash}p{1cm}>{\centering\arraybackslash}p{1cm}|>{\centering\arraybackslash}p{1cm}>{\centering\arraybackslash}p{1cm}|}
\cline{2-9}
\multicolumn{1}{l|}{} & \multicolumn{2}{c|}{$\frank$} & \multicolumn{2}{c|}{$\brank$} & \multicolumn{2}{c|}{$\hrank$} & \multicolumn{2}{c|}{$\urank$} \\
&~$r_{s}$&~$r_{\tau}$&~$r_{s}$&~$r_{\tau}$&~$r_{s}$&~$r_{\tau}$&~$r_{s}$&~$r_{\tau}$\\
\cline{2-9}
& \multicolumn{8}{c|}{a) Maritime}\\
\cline{1-9}
\multicolumn{1}{|l|}{$\frank$}  & - & - & $0.96$ & $0.85$ & $0.95$ & $0.81$ & $0.95$ & $0.81$\\
\multicolumn{1}{|l|}{$\brank$} & - & - & - & -  & $0.98$ & $0.89$ & $0.90$ & $0.74$\\
\multicolumn{1}{|l|}{$\vrank$} & $0.94$ & 0.$81$ & $0.99$ & $0.92$  & $0.97$ & $0.85$ & $0.89$ & $0.71$\\
\cline{1-9}
& \multicolumn{8}{c|}{b) Airports}\\
\cline{1-9}
\multicolumn{1}{|l|}{$\frank$} & - & - & $0.98$ & $0.91$ & $0.96$ & $0.86$ & $0.95$ & $0.83$ \\
\multicolumn{1}{|l|}{$\brank$} & - & - & - & -  & $0.99$ & $0.91$ & $0.91$ & $0.79$ \\
\multicolumn{1}{|l|}{$\vrank$} & $0.98$ & $0.91$ & $0.998$ & $0.96$  & 0.99 & $0.92$ & $0.92$ & $0.80$\\
\cline{1-9}
& \multicolumn{8}{c|}{c) Taxis}\\
\cline{1-9}
\multicolumn{1}{|l|}{$\frank$} & - & - & $0.62$ & $0.48$ & $0.44$ & $0.34$ & $0.77$ & $0.61$ \\
\multicolumn{1}{|l|}{$\brank$}& - & - & - & -  & $0.94$ & $0.82$ & $0.88$ & $0.70$ \\
\multicolumn{1}{|l|}{$\vrank$}  & $0.42$ & $0.30$ & $0.92$ & $0.79$ & $0.98$ & $0.91$ & $0.76$ & $0.58$\\
\cline{1-9}
\end{tabular}
}
\end{table}

The $\nrep$-bias also affects the Top 10s ranking which is a popular usage of PR ranking.
The Top 10s 
related to Maritime are displayed in Table~\ref{tab:top10_maritime}. 
Ports with bold name are new entries when compared to the previous ranking.
Although $80\%$ of entries are common to all Top 10s, the differences come from reordering. Both $\hrank$ and $\brank$ fit almost perfectly with the ten most represented ports ($\reprank$). On the other hand, $\urank$ may capture items with bad $\reprank$ \textit{e.g.} the port of Surabaya  ($\reprank = 45$ and $\urank = 9$).
\begin{table}[h!]
\centering
\caption{
    Maritime's Top 10s PageRank.}
\label{tab:top10_maritime}
\resizebox{\textwidth}{!}{
    \begin{tabular}{|l|lcc|lcc|lcc|lcc|}
\cline{2-13}
\multicolumn{1}{l|}{}  & \multicolumn{3}{c|}{$\frank$} & \multicolumn{3}{c|}{$\brank$} & \multicolumn{3}{c|}{$\hrank$} & \multicolumn{3}{c|}{$\urank$}\\
\cline{1-13}
Rank & Port ~& $\reprank$ ~& $\vrank$~ & Port ~& $\reprank$ ~& $\vrank$ & Port ~& $\reprank$ ~& $\vrank$~ & Port ~& $\reprank$ ~& $\vrank$\\
\cline{1-13}
1&Singapore&2&2~&Singapore&2&2&Hong Kong  &1&1~&Singapore  &2&2\\
2&Hong Kong  &1&1~&Hong Kong  &1&1&Singapore  &2&2~&Busan  &4&4\\
3&Rotterdam  &5&7~&Shanghai  &3&3&Shanghai  &3&3~&Hong Kong  &1&1\\
4&Busan  &4&4~&Busan  &4&4&Busan  &4&4~&Rotterdam  &5&7\\
5&Shanghai  &3&3~&Rotterdam  &5&7&Rotterdam  &5&7~&Shanghai  &3&3\\
6&Hamburg  &8&10~&Port Klang  &6&6&Port Klang  &6&6~&Hamburg  &8&10\\
7&Port Klang  &6&6~&Kaohsiung  &7&5&Kaohsiung  &7&5~&Antwerp  &10&12\\
8&Antwerp  &10&12~&Hamburg  &8&10&Hamburg  &8&10~&\textbf{Bremerhaven}  &12&19\\
9&Bremerhaven  &12&19~&Antwerp  &10&12&Antwerp  &10&12~&\textbf{Surabaya}  &45&36\\
10&Kaohsiung  &7&5~&\textbf{Jebel Ali}  &9&11&Jebel Ali  &9&11~&Port Klang  &6&6\\
\cline{1-13}
\end{tabular}
}
\end{table}

\begin{figure}[h!]
    \centering
    \begin{subfigure}[t]{0.23\columnwidth}
        \centering
        \includegraphics[width=\textwidth]{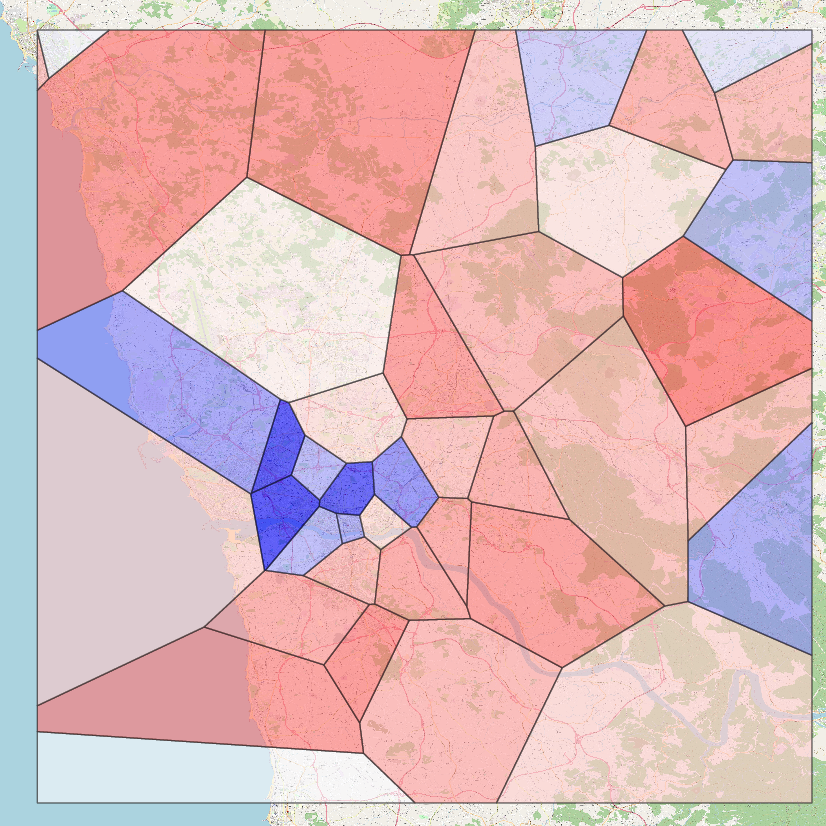}
        \caption{$\fprvec$}
        \label{fig:portomap_f}
    \end{subfigure}
    \begin{subfigure}[t]{0.23\columnwidth}
        \centering
        \includegraphics[width=\columnwidth]{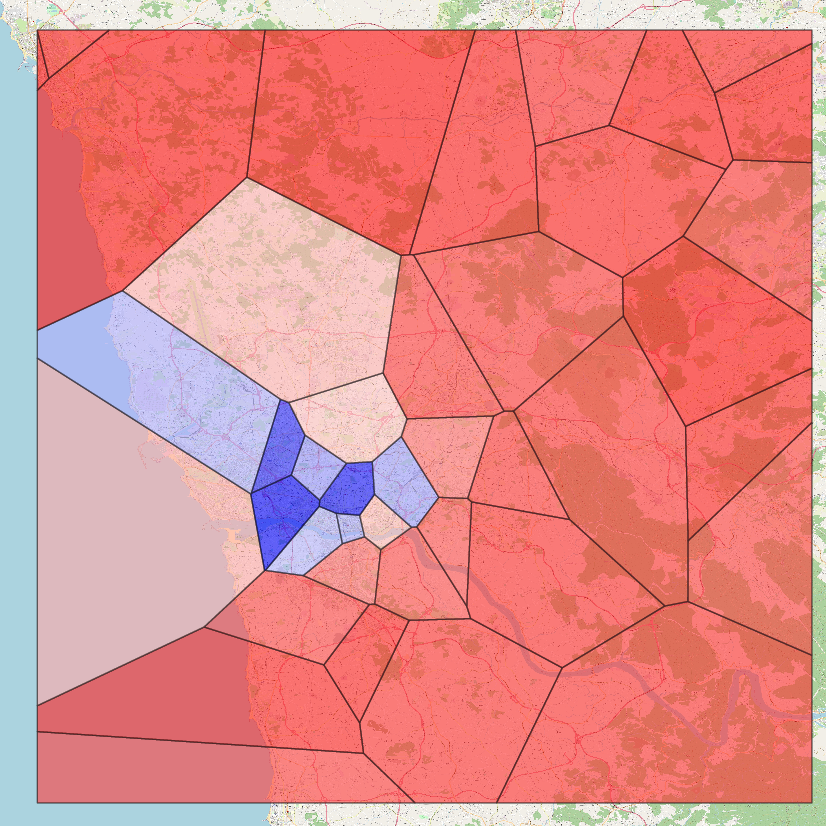}
        \caption{$\bprvec$}
        \label{fig:portomap_b}
    \end{subfigure}
    \begin{subfigure}[t]{0.23\columnwidth}
        \centering
        \includegraphics[width=\columnwidth]{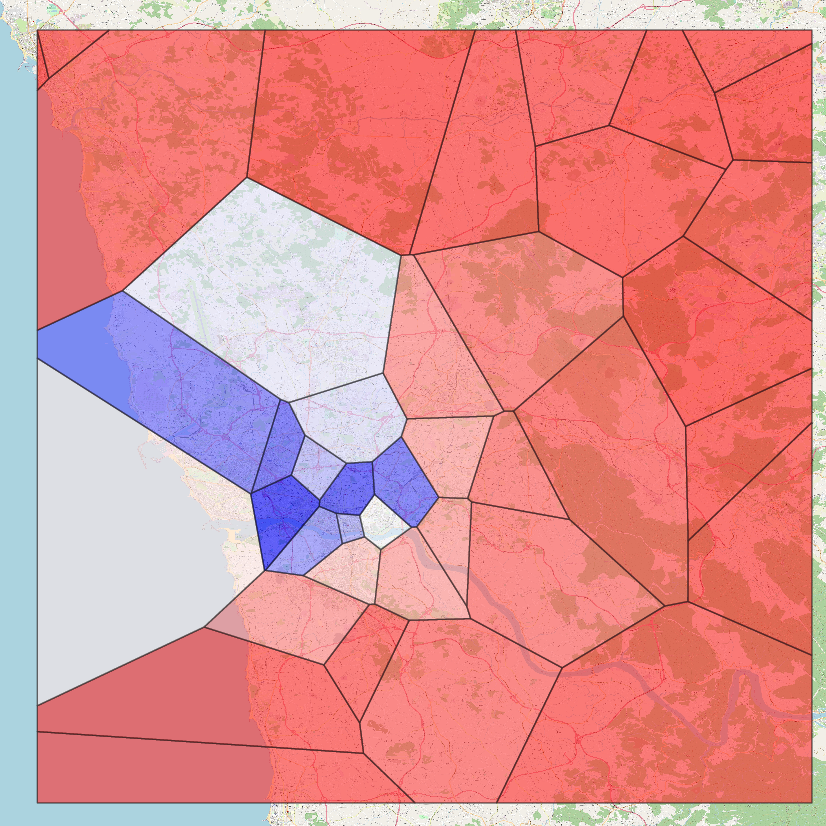}
        \caption{$\hprvec$}
        \label{fig:portomap_h}
    \end{subfigure}
    \begin{subfigure}[t]{0.23\columnwidth}
        \centering
        \includegraphics[width=\columnwidth]{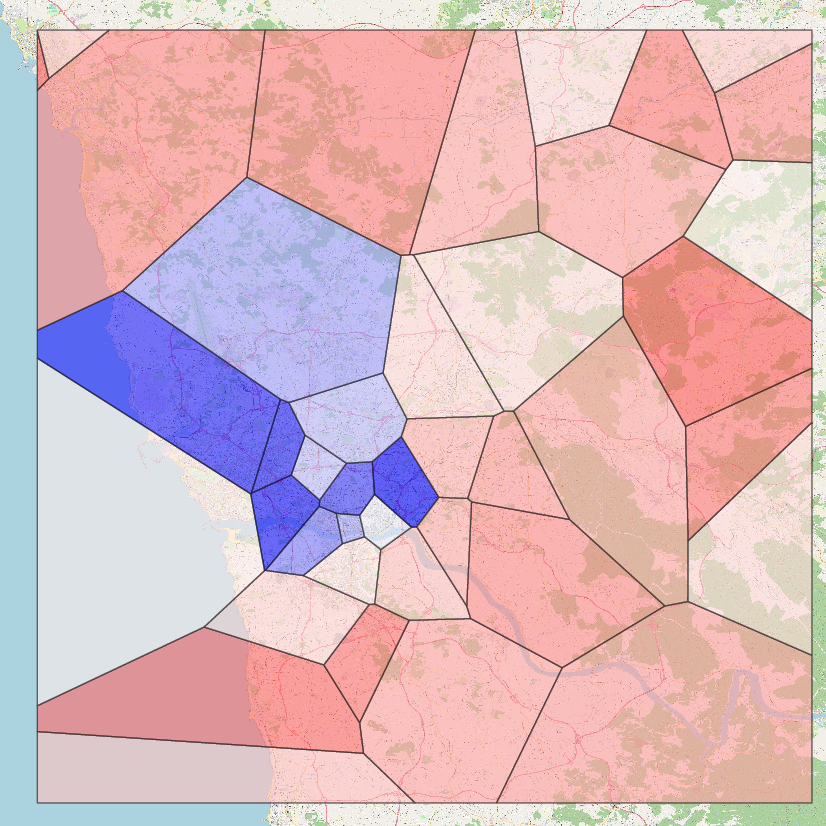}
        \caption{$\uprvec$}
        \label{fig:portomap_u}
    \end{subfigure}
    \includegraphics[width=0.35\columnwidth]{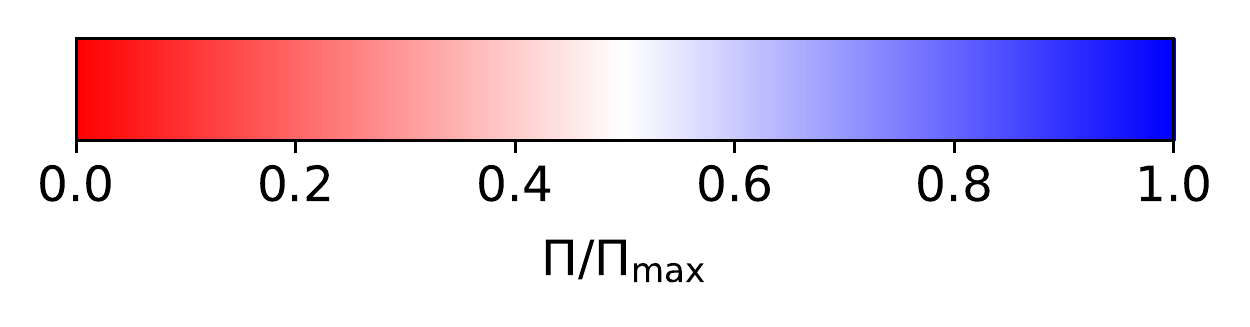}

    \caption{Distribution of PageRanks values for Porto's neighborhood.}
    \label{fig:portomap}
\end{figure}

Since the number of items composing Taxis dataset (corresponding to subareas of Porto) is small enough, the PR scores of all items are given in Fig.~\ref{fig:portomap}. Both $\hprvec$ and $\bprvec$ give bad ranks to peripherals. Only {\fston}PR and Unbiased {\von}PR models give importance to peripheral neighborhoods. Finally, central regions have similar rankings whatever the model used.


\begin{figure}[h!]
\centering
\begin{subfigure}[b]{0.3\textwidth}
\centering
\includegraphics[width=\textwidth]{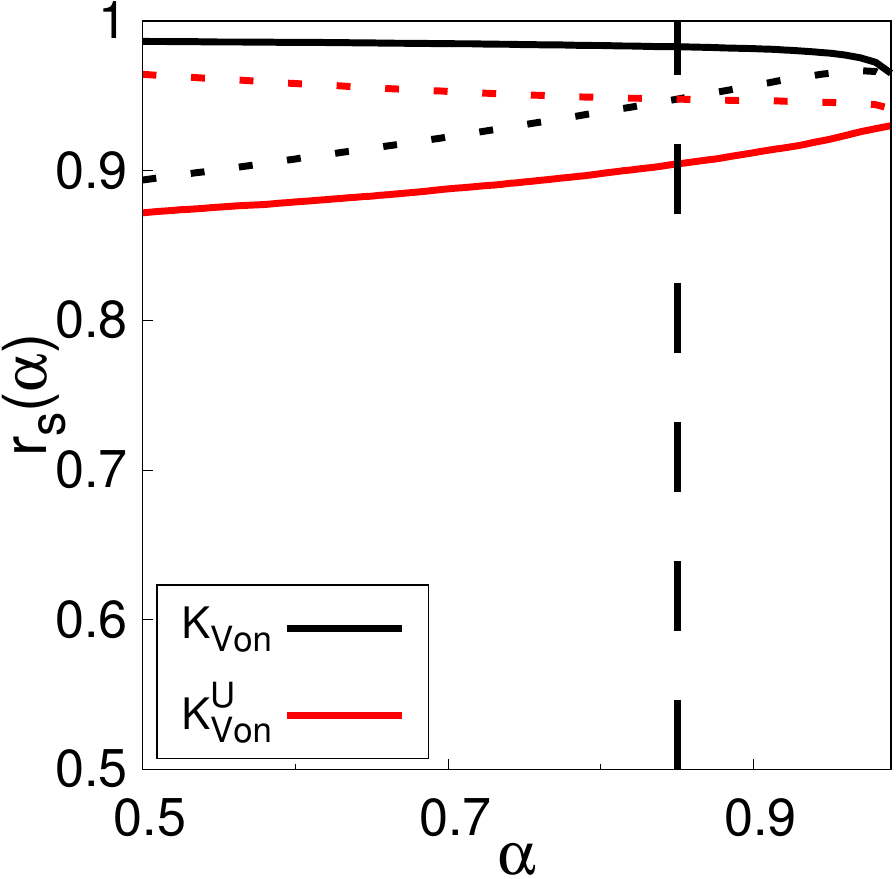}
\caption{Maritime}
\label{fig:alpha_maritime}
\end{subfigure}
\begin{subfigure}[b]{0.3\textwidth}
\centering
\includegraphics[width=\textwidth]{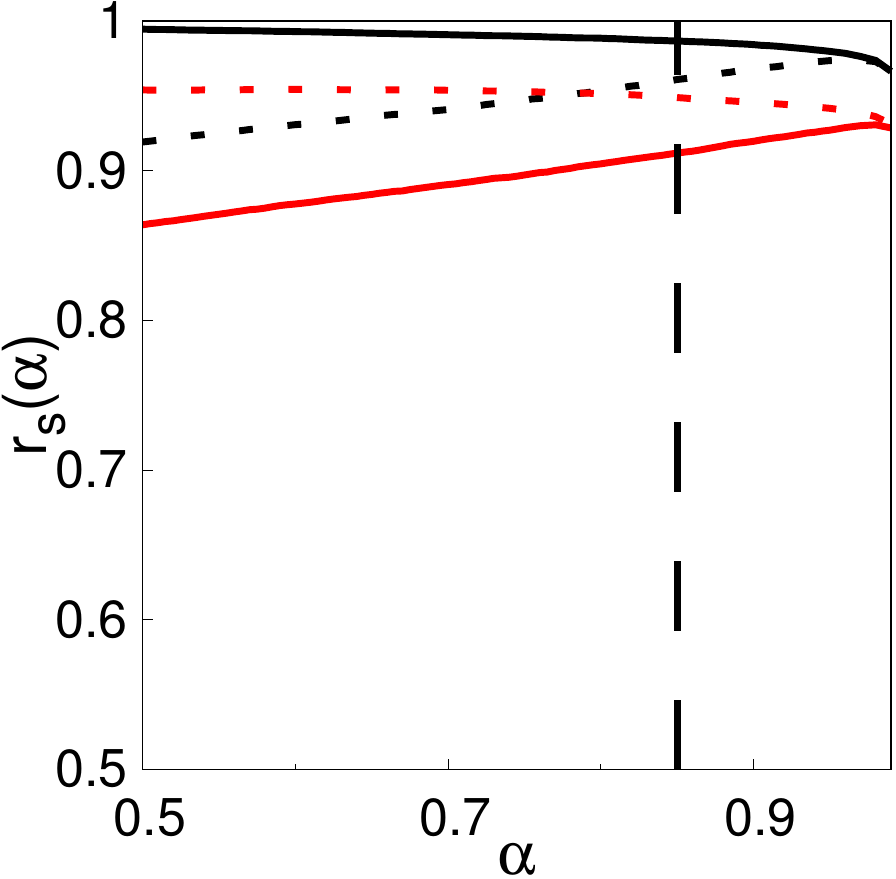}
\caption{Airports}
\label{fig:alpha_airports}
\end{subfigure}
\begin{subfigure}[b]{0.3\textwidth}
\centering
\includegraphics[width=\textwidth]{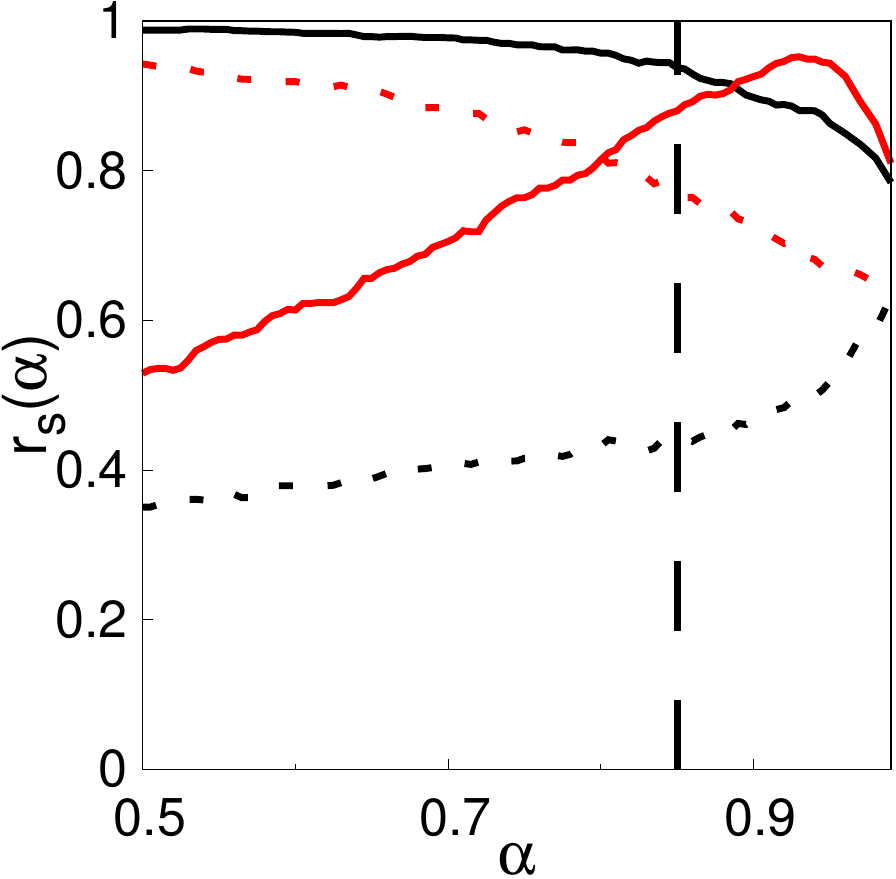}
\caption{Taxis}
\label{fig:alpha_taxis}
\end{subfigure}
\caption{Evolution of the Spearman correlation coefficient $r_{s}(\alpha)$ with $\alpha$, for couples of rankings. Solid lines (dashed lines) are related to correlations with Biased {\fston}PR ({\fston}PR). Vertical black dashed line represents $\alpha=0.85$.}
\label{fig:alpha}
\end{figure}

\paragraph{Dependence of the $\nrep$-bias with the damping factor $\alpha$.}
Since in the literature an alternative value of the damping factor $\alpha$ could be used (as in \cite{coquide19b}), we investigated similarities between rankings regarding its choice.
The evolution of Spearman correlation coefficient with respect to changes in $\alpha$ is present in Fig.~\ref{fig:alpha} for $\alpha \in [0.5, 0.99]$. Results related to the Kendall correlation coefficient evolution are not reported since they are similar. 
For $\alpha \leq 0.85$, the observations made earlier are still valid. When teleportations are less frequent, different changes occur. 
Indeed, for Maritime and Airports, $\hrank$ becomes closer to $\frank$ than $\urank$ is to $\frank$ (dashed lines).
Overall the pairs $\hrank$-$\urank$ and $\frank$-$\brank$ get closer as $\alpha$ tends to $1$ due to the poor contribution of teleportations. 
For the taxis, we notice a switch at $\alpha \approx 0.9$ for $\brank$ (solid lines). We think this is due to the low amount of items related to Taxis dataset. In order to understand this behaviour, we need to further investigate other similar datasets. 

\section{Future works}
\label{sec:future_works}

This study shows that the application of network measures to the new objects that are  \textsc{Von}s are not trivial.
We believe the adaptation of other analysis tools is an important challenge for the network science community.
We are currently investigating the application of clustering algorithms such as Infomap~\cite{rosvall09} to \textsc{Von}s.
This algorithm indeed uses PR in order to compare clustering qualities. However, \cite{xu16} also suggests that such algorithm can be directly applied to {\von}s with no modifications.
The PR centrality measure has other applications.
A recent method based on the Google matrix (the stochastic matrix which models the random surfer), called \textit{reduced Google matrix}, has shown its efficiency in inferring hidden links between a set of nodes of interests \cite{frahm16b} for example with studying Wikipedia networks~\cite{coquide19a}.
Using user traces on website rather than usual hypertext click statistics, we will also study the generalization of this tool to {\von}s.

%
\bibliographystyle{splncs03}
\bibliography{draft}

\end{document}